# Breaking the Legend: Maxmin Fairness notion is no longer effective

Yaser Miaji *MIEEE*
InterNetWorks Research Group
UUM College of Arts and Sciences, University Utara Malaysia, 06010 – UUM Sintok, Malaysia
S92171@student.uum.edu.my
Suhaidi Hassan PhD *SMIEEE*
InterNetWorks Research Group
UUM College of Arts and Sciences, University Utara Malaysia, 06010 – UUM Sintok, Malaysia
Suhaidi@ieee.org

ABSTRACT

*In this paper we analytically propose an alternative approach to achieve better fairness in scheduling mechanisms which could provide better quality of service particularly for real time application. Our proposal oppose the allocation of the bandwidth which adopted by all previous scheduling mechanism. It rather adopt the opposition approach be proposing the notion of Maxmin-charge which fairly distribute the congestion. Furthermore, analytical proposition of novel mechanism named as Just Queueing is been demonstrated.*

## I. *INTRODUCTION*

Since the recognition and discovery of the congestion in packet switching network by Nagle, many scholars attempts to reach a respective solution to remedy such issue. The most appreciated mechanism is Weighted Fair Queueing WFQ which proposed by Demers in 1998 [1]. Since the success of WFQ, most proceeding scheduling Time Stamp (TS) scheduling algorithm proposed in the literature are based on the principle of maxmin which first proposed by Demers.

However, by the rapid emergence and increase popularity of much loss and delay sensitive application such as online-gaming, Skype, IPTV and video conferencing, network re-suffer from the same problem from different prospective [2]. The engagement of greedy programs such as download manager leads to ideality of the network from the user of such program, whereas other users complain from unfairness and poor Quality of Service (QoS).

Since most scheduling mechanisms utilized in today's routers and most proposed mechanism in the literature highly depend in two main principles namely; fair allocation and abandoning source as user definition in their mechanisms, network will suffer more in the future.

Therefore, this article designed to steer scholars' attention to neglected direction based on the theory of justice. Just Queueing (JQ) is conceptually proposed in this article which based in Rawls' theory of justice. Future papers will illustrate the mathematical and simulation approaches.

There are three fundamental aims of this article. Firstly, it introduces theory of justice and engages such theory in information technology science. Secondly, it injects the principles of theory of justice in scheduling mechanisms. Finally, it proposes JQ as an enhancement mechanism for WFQ.

Nest section thoroughly studies the principles of the theory of justice. It follows by the injection of the principles in the scheduling mechanisms. Fourth section presents a comparison between WFQ principles and theory of justice principles. Finally, JQ is conceptually, proposed.

The fabulous weighted fair queueing (WFQ [3] accosts the issue of implementing scheduling properties with appropriate level of adequacy and comprehensive coverage in term of explanation and simulation. The principle is to allocate the bandwidth among the users fairly according to the rule of maximizing the flow that utilizes the bandwidth least. The consequence of this approach is significant in research community. In fact, it inspires most, if not all, the significant proceeding schedulers such as WF2Q, WF2Q+, SFQ, SCFQ, SPFQ, VC, Delay-DDR, Jitter-DDR, WRR, DRR, MDRR, SRR and so forth, particularly in the principle of fairness in bandwidth allocation [3-25]n.

This article will propose an alternative approach which could be considered as an opposition to the WFQ's bandwidth fairness allocation principle. Our approach will be posed conceptually. The mathematical and simulation approaches could be introduced in future papers. As it has been stated early, this paper introduces an alternative approach based on justice theory and will be discussed by conceptual comparison and contrast with the bandwidth allocation in WFQ.

There are three main goals of this article. Firstly, this article will introduce justice theory as a fundamental theory for designing a scheduler specially to provide fairness in scheduling algorithm. The second objective is to inject the principle of justice theory in the scheduling algorithm. The conceptually comparison between WFQ and just queueing (JQ) will be the last objective.

Next section presents the principle of theory of justice and the motivation of its involvement in the scheduling algorithm. This will be followed by the synthesis between the theory of justice and the scheduling algorithm. The demonstration of the WFQ concept will be next stage as an introductory stage for the final stage which includes the contrast and the comparison.

## *II. THEORY OF JUSTICE*

Theory of Justice [26] is a social theory which could be used in several scientific fields. This theory has two main principles; liberty and difference. These principles could be incorporated in the scheduling mechanism. Therefore, this section will assist in this incorporation. Firstly, major principles of the theory will be illustrated and then the incorporation will be evolved gradually.

### 2.1 Rawls' Theory

Theory of justice is a globally discussed theory of political and moral philosophy first initiated by John Rawls. In A Theory of Justice, Rawls attempts to work out the issue of distributive justice by utilizing a variant of the familiar device of the social contract. The resulting theory is known as "Justice as Fairness" [27], from which Rawls develop his two famous principles of justice: the liberty principle and the difference principle. According to Rawls, ignorance of these details about oneself will lead to principles which are fair to all. If an individual does not know how he will end up in his own conceived society, he is likely not going to privilege any one class of people, but rather develop a scheme of justice that consider all fairly. In particular, Rawls asserts that those in the Original Position would all adopt a maximin strategy which would maximize the position of the least well-off. Rawls seeks to persuade us through argument that the principles of justice that he derives are in fact what we would agree upon if we were in the hypothetical situation of the original position and that those principles have moral weight as a result of that. It is non-historical in the sense that it is not supposed that the agreement has ever, or indeed could actually be entered into as a matter of fact. Rawls claims that the parties in the original position would adopt two such principles, which would then govern the assignment of rights and duties and regulate the distribution of social and economic advantages across society.

### 2.2 Principles of Justice

#### 2.1.1 Right Principle

- Each person is to have an equal right to the most extensive total system of equal basic liberties compatible with a similar system of liberty for all

#### 2.1.2 Elucidation of Right Principle

The basic liberties of citizens are, roughly speaking, political liberty (i.e., to vote and run for office); freedom of speech and assembly, liberty of conscience and freedom of thought, freedom of the person along with the right to hold (personal) property; and freedom from arbitrary arrest. It is a matter of some debate whether freedom of contract can be inferred as being included among these basic liberties. The first principle is more or less absolute, and may not be violated, even for the sake of the second principle, above an unspecified but low level of economic development (i.e. the first principle is, under most conditions, lexically prior to the second principle). However, because various basic liberties may conflict, it may be necessary to trade them off against each other for the sake of obtaining the largest possible system of rights. There is thus some uncertainty as to exactly what is mandated by the principle, and it is possible that a plurality of sets of liberties satisfy its requirements.

#### 2.1.3 Good Principle

- Social and economic inequalities are to be arranged so that they are both :

  (a) To the greatest benefit of the least advantaged, consistent with the just savings principle, and

  (b) Attached to offices and positions open to all under conditions of fair equality of opportunity.

#### 2.1.4 Elucidation of Good Principle

Rawls' claim in b) is that departures from equality of a list of what he calls primary goods – 'things which a rational man wants whatever else he wants' are justified only to the extent that they improve the lot of those who are worst-off under that distribution in comparison with the previous, equal, distribution. His position is at least in some sense egalitarian, with a proviso that equality is not to be achieved by worsening the position of the better-off. An important consequence here, however, are those inequalities can actually be just on Rawls's view, as long as they are to the benefit of the least well off. His argument for this position rests heavily on the claim that morally arbitrary factors (for example, the family we are born into) should not determine our life chances or opportunities. Rawls is also keying on an intuition that we do not deserve inborn talents, thus we are not entitled to all the benefits we could possibly receive from them, meaning that at least one of the criteria which could provide an alternative to equality in assessing the justice of distributions is eliminated. The stipulation in b) is prior to that in a) and requires more than meritocracy. 'Fair equality of opportunity' requires not merely that offices and positions are distributed on the basis of merit, but that all have reasonable opportunity to acquire the skills on the basis of which merit is assessed. It is often thought that this stipulation, and even the first principle of justice, may require greater equality than the difference principle, because large social and economic inequalities, even when they are to the advantage of the worst-off, will tend to seriously undermine the value of the political liberties and any measures towards fair equality of opportunity.

#### 2.1.5 Maxmin Rule

- rank alternatives by the worst possible outcome (you can belong to the lowest/poorest group in the real society)

#### 2.1.6 Elucidation

Rawls claims that people use the maximin rule to choose principles of justice in his original position. According to the maximin rule we should compare alternatives by the worst possible outcome under each alternative, and we should choose one which maximizes the utility of the worst outcome.

This rule is rational under certain conditions. First, we do not know the probability of each circumstance under each decision. This makes it impossible to calculate expectation of gain. Second, the worst off position chosen by maximin rule is good enough that we are not eager to get more than that. Third, the worst positions under other alternatives are unacceptably bad. Under the second and third assumptions we are inclined to secure the minimal acceptable result above all. Thus we use the maximin rule. Rawls thinks that original position satisfies these conditions.

As for the first assumption, we do not know anything about the probability of each outcome by the veil of ignorance. We do not know even exact values of outcomes. To consider second and third assumption, we should recall the alternatives among which we are choosing. We restricted the alternatives to the theories which are actually proposed as serious theories. Among them, most dominant ones are classical and average utilitarianism or usefulness, and two principles of justice. Among these alternatives, the two principles of justice assure us a minimal acceptable outcome, due to the first principle and the difference principle (And under the condition of moderate scarcity, the worst off position under this alternative is acceptable by the definition of "moderate"). We can also assume the law of diminishing marginal utility as a psychological fact. According to this law, when we have already had enough, adding more goods does not increase our utility a lot. Thus we are not eager to get more than minimal enough outcomes. Third assumption is understandable if we think about other dominant alternatives, namely classical and average utilitarianism. Under these alternatives, someone can be unacceptably worse off for the sake of maximizing utility. Utilitarian's say this cannot happen because of diminishing marginal utility, but this is only a conjecture or a guess. We cannot risk the secure minimal outcome by such a conjecture. Therefore, the second and third assumptions are satisfied in the original position.

## *III. WEIGHTED FAIR QUEUEING*

The necessity of clarification of WFQ is obvious since the proposed mechanism is an enhancement of WFQ. This section shall cover the conceptual and mathematical approach of WFQ.

Maxmin fairness concept is first adopted by WFQ [3] adopted and was defined according to the fairness principles which are explained and justified next. Firstly, the maximum will not be exceeded. Secondly, guarantee of the stability of the minimum allocation to a flow in all situation. Finally, Accomplishment of minimum user request according to first and second principle will increase the total resources. There are two major drawbacks have been deduced from the previous maxmin conditions. At first, if the flow associated with the source-destination principle, the vulnerability of the network is high since malicious users could establish several sessions with different destination and hence gain larger bandwidth. Secondly, network breakthrough could be achieved by establishing multiple small packet session by the greedy host which will also subdue the maxmin barrier. Even though, WFQ is the best mechanism, so far, which fulfills the requirements of the scheduler with acceptable level, scholars are looking forward to enhance such mechanism which published in 1989.

Furthermore, WFQ adopted the principle of fair allocation of the bandwidth. Its approach primarily attempted to approximate the generalized process sharing (GPS), which impossible to be exactly reached, since it proposes the infinitesimal. Hence, the approximation accomplished with more than one packet difference in each flow. The case could be worse, according to [4] and [5], in the presence of the congestion. Furthermore, most of the proceeding scheduler, despite of their difference in methodology and algorithm, have the same major principle which is the fair allocation of the bandwidth among the flows, even though, the flows have been identified differently. Therefore, the concept remains stable with a minor deviation on some other sub-concepts.

As a result, WFQ based its mathematical equation on the previous mentioned principles and the outcome are presented in this section. In WFQ, as the packet arrive in a specific queue according to a specific criterion defined by the scheduling algorithm, its timestamp calculated according to its virtual arrival time and virtual finish time of the previous packet in the same queue. As a consequence, its virtual finish time is set as presented in Eq (1-a) and (1-b):

$$S_i^n = \max(F_i^{n-1}, V(a_i^n)) \dots\dots\dots\dots\dots\dots\dots\dots\dots 1-a$$

$$F_i^n = S_i^n + \frac{L_i^n}{\emptyset_i} \dots\dots\dots\dots\dots\dots\dots\dots\dots 1-b$$

Where,

S = Virtual Start time

F = Virtual finish time for the previous packet

A = Virtual arrival time

L = packet length

Eq. (1-a) and (1-b) is applied for all active flow either in congestion situation or in non-congestion case. The case in JQ is quite different as it is presented in next section.

## *IV. JUST QUEUEING MECHANISM*

This section will analytically discuss the proposed modification of the scheduling in packet switch network. Firstly, the incorporation of theory of justice in the networking and specifically, in scheduling mechanism will be explained. The next section will define the proprieties of the new mechanisms and define some new terminology.

The principle of maxmin has been explained in Rawl's theory as "*rank alternatives by the worst possible outcome (you can belong to the lowest/poorest group in the real society)"* [1]. Consequently, from the above definition, reader could infer that this principle is applicable either in the distribution of the available resources or the distribution of the lack of resources. Before the discussion exemplified, there will be an engagement of other principles; the principle of rights and the principle of good. Rawls elaborate the *rights* as a concept applied to action and circumstances in accordance with principles which means that a rational person concerned to advance his interests would accept in a position of equality to settle the basic term of his association. The next principle is the definition of *good*; if an object has the properties that it is rational for someone with rational plan of life to wont, and then it is good for him [27].

For example, if the fund considered as a resource. Hence, whenever there is a fund available in the institution's resources, the administration has the right to distribute this fund according to a principle such as the person position or technical demand or whatever. Contrary, if this fund is considered to be extra then the distribution could be based on the demand according to the rational plan of the department which could improve the ability. Therefore, the fairness and justices have determined according to the purpose of the distribution or the use and the equality principle.

Another example which illustrates the opponent, if the case is to distribute the charge then the actions and the results will be different. So, if the institute involved in several projects which could result in heavy tasks, then the charge of late job could be distributed among the departments according to the same principles of rights and good. Consequently, depend on the adopted principle such as the punctuality principle, the department, which insufficiently accomplish the task in a specific timeframe, will be charged according to the accumulative time wastage. Another approach which depends on the good principle, it is good to charge this department which got sufficient fund from previous tasks and so forth. The above discussion on the distribution of the resources and the charge is also applicable for the Internet analogy which will be elaborated later.

Referring to the analogy of resource allocation and *charge allocation*, all of the previous scheduler attempt to fairly allocate the bandwidth among the users or the flows or whatever algorithm they adopted which is similar to the example of fund allocation. By comparison, JQ proposes different approach which adopts the principle of fair allocation of the charges on other words fair allocation of the congestion.

*Fair charge* could improve the scheduling algorithm in different ways. The vulnerability of the scheduler to the malicious user could be reduced or may be eliminated by imposing and allocating part of the congestion effect on this specific user. Also, there could be an involvement of rights and good principle. The flow which consumes much bandwidth could allocate much congestion or in other words could suffer more in occurrence of the congestion; however, this will not affect the network performance since its queue could be less than the others.

Every flow has its good and right. For instance, it is the rights of audio and video packets to have a priority in the transmission since they are delay sensitive. Furthermore, it is good for FTP to be transmitted earlier (if there are no rights for others). So, hereby, we declare two novel definitions: the right flow and the good flow. These two definitions will be integrated into the equation of JQ.

The concept is basically, the network will behave normally in the absence of the congestion. As the congestion occurs or expected, the algorithm is to impose some charges upon those how are highly expected to be the cause of the congestion. Also, at this time the network starts to rearrange its mechanism in order to identify the prioritization of the congestion distribution. Consequently, the misbehaved user will be charged more and the network will be stable again. Hence, from the above discussion, there are three novel elements introduced in JQ namely; fair charge allocation which replaces fair resource allocation, right flow and good flow. Next section explains who these principles integrated into the equation of JQ.

Therefore, in the absent of congestion problem the network will behave according to Eq. (1-a) and (1-b). However, as the congestion appears or expected the equation will slightly change to incorporate fair charge, right flow and good flow as it is presented in Eq. (2-a) and (2-b):

$$S_i^n = max\,(F_i^{n-1}, V(a_i^n)) + fl_{le} \ldots\ldots\ldots\ldots\ldots\ldots\ldots 2-a$$

$$F_i^n = S_i^n + \frac{L_i^n + Usr_{le}}{\emptyset_i} \ldots\ldots\ldots\ldots\ldots\ldots\ldots\ldots\ldots 2-b$$

Where:

$fl_{le}$ = is the level of the flow, For example voice packet is assigned level 1 since it is delay sensitive whereas Simple Mail Transfer Protocol (SMTP) could be assigned level 4 since it has no sensitivity for the delay. The flow level is calculated according to ToS field in IPv4 or pri in IPv6.

$Usr_{le}$ = is the user level. For instance user who is suspected to be a greedy will be assigned higher level, therefore, its virtual finish time could be higher and hence its transmission is delayed.

## V. ANALYSIS

The analysis procedure is by comparing and contrasting Just Queueing (JQ) and WFQ as it is the most adopted scheduler which carries similar discipline to other scheduling mechanisms. As it has been avowed earlier in the justice theory section, the definition of fairness in theory of justice is closely related to the definition of fairness which has been adopted by WFQ with some difference. WFQ tries to approximate the GPS which fairly allocate the resource with infinite storage capability. The maxmin principle is quite similar to the justice theory maxmin. However, the subtle difference is the lack of the incorporation of the good and right principle.

In justice theory, even with the presence of the minimum, there should be a consideration of the right and good. So, referring to the three principle of maxmin which identified in WFQ, justice theory differ in two more which are; the rights and good. Consequently, the distribution of the resource according to the justice theory should involve the consideration of right of using this resource according to a principle and the consideration of the good according to the plan of the packet or stream of packets.

The proposed JQ is an enhancement for WFQ in the congestion case. It integrates the principle of $Usr_{le}$and $fl_{le}$ for defining the user malicious attack and the sensitivity of the packet respectively. Therefore, at the normal implementation with no sign of congestion, the mechanism will behave typically like WFQ. The behavior changes as the congestion occurs or is expected.

To conclude, JQ and WFQ share some similarities as well as differences. The similarities are in the fairness definition. However, there should be an incorporation of the rights and good principles as well.

## VI. SUMMARY

The enhancement for Weighted Fair Queueing (WFQ) mechanism has been proposed in this paper mathematically and conceptually based on the theory of justice by Rawls. The WFQ fairness principle has been demonstrated. This followed by some remarkable similarities and differences between justice theory on a hand and WFQ, as the representative of scheduler, on another. Finally, there has been a proposition of enhanced algorithm which conceptually different from WFQ. Also, this proposition is conceptually and mathematically elaborated and named as Just Queueing (JQ), therefore, simulation approach will be presented in future papers. It also, introduce three novel concepts namely: fair charge, good flow and right flow. Therefore, the calculation of the start and finish time for WFQ at the presence of congestion is slightly different. The main aim of such mechanism is to prevent the network from the misbehaved users. This could provide sustainability to the network. It also, supports the Quality of Service (QoS) by allowing the delay sensitive packet to be transmitted first. Therefore, JQ provide more protection and better QoS.

**Authors**

**Eng. Yaser Miaji**

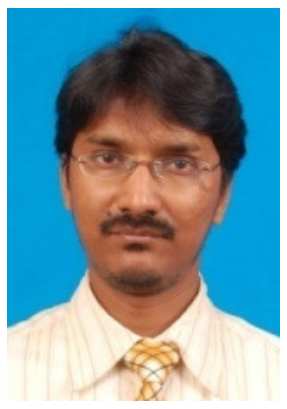

Doctoral researcher in Information Technology College of Arts and Sciences, University Utara Malaysia

Master in Telecommunication Engineering University of New South Wales, Australia, 2007

Bachelor in Electrical Engineering, Technical College in Riyadh, Saudi Arabia, 1996

Lecturer in College Of Telecommunication and Electronics in Jeddah,

Member of IEEE and ACM

**Dr. Suhaidi Hassan**

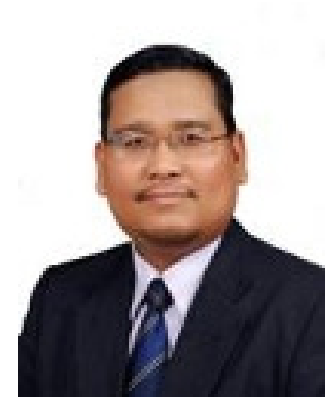

Associate Professor & Assistant Vice Chancellor at University Utara Malaysia

PhD degree in Computing (specializing in Networks Performance Engineering) from the University of Leeds in the United Kingdom.

MS degree in Information Science (with concentration in Telecommunications and Networks) from the university of Pittshugh, USA.

BSc degree in Computer Science from Binghamton University, USA.

Senior Member of IEEE